\documentclass[conference]{IEEEtran}
\IEEEoverridecommandlockouts

\usepackage[T1]{fontenc}
\usepackage{graphicx}
\usepackage{amsmath}
\usepackage{amssymb}
\usepackage{textcomp}
\usepackage{xcolor}
\usepackage{booktabs}
\usepackage{enumitem}
\usepackage{multirow}
\usepackage{array}
\usepackage{pifont}
\newcommand{\cmark}{\ding{51}}
\usepackage{tikz}
\usetikzlibrary{arrows.meta,positioning,fit,backgrounds,shapes.geometric}
\usepackage{hyperref}
\usepackage[capitalise,noabbrev]{cleveref}

\newcommand{\system}{\textsc{ToolMinimize}}
\newcommand{\pc}{\text{PC}}

\begin{document}

\title{ToolMinimize: Auditing and Rewriting\\
       LLM Agent Tool Calls to Minimize Privacy Exposure}

\author{
\IEEEauthorblockN{Wenbiao Li}
\IEEEauthorblockA{\textit{Case Western Reserve University}\\
Cleveland, OH, USA\\
wxl387@case.edu}
\and
\IEEEauthorblockN{Yuqiao Xu}
\IEEEauthorblockA{\textit{Case Western Reserve University}\\
Cleveland, OH, USA\\
yxx914@case.edu}
\thanks{Accepted for publication in the Proceedings of the 23rd Annual
International Conference on Privacy, Security and Trust (PST 2026),
University of Ottawa, Ottawa, Canada, August 26--28, 2026. This is the
authors' accepted version; the final published version will appear in
IEEE~Xplore (DOI to be added upon publication).}
\thanks{\copyright~2026 IEEE. Personal use of this material is permitted.
Permission from IEEE must be obtained for all other uses, in any current or
future media, including reprinting/republishing this material for advertising
or promotional purposes, creating new collective works, for resale or
redistribution to servers or lists, or reuse of any copyrighted component of
this work in other works.}
}

\maketitle

\IEEEpubid{\makebox[\columnwidth]{\copyright2026 IEEE\hfill}\hspace{\columnsep}\makebox[\columnwidth]{ }}

\begin{abstract}
LLM agents routinely include privacy-sensitive data (PSD) in tool call arguments beyond what the invoked tools require, crossing trust boundaries to third-party services on every invocation. A controlled measurement on three production LLMs (GPT-4o, Claude 3.5 Sonnet, Llama-3.3-70B) shows that 81--88\% of tool calls include unnecessary PSD under default prompts; explicit privacy instructions still leave 36--76\% over-sharing.  Existing defenses gate calls (allow/block) or label flows (information-flow control) but cannot \emph{rewrite} argument values, and PII detection tools miss implicit PSD like ``Memorial Sloan Kettering'' (a hospital name that implies a diagnosis).  We present \system{}, a middleware that intercepts tool calls and rewrites their arguments to the minimum data necessary for tool functionality, combining schema-aware necessity analysis with four operations: removal, generalization, substitution, and truncation.  Live validation on 307 tool calls across the three LLMs above reduces privacy cost by 81.2--92.0\% at 100\% argument-level task validity (TOST equivalence $p{<}0.001$ at $\Delta{=}1.0$); on 25 unannotated Model Context Protocol (MCP) schemas, by 79.0\% with no \texttt{minimum\_necessary} metadata.  An optional LLM content-necessity layer strips task-irrelevant PSD from otherwise-necessary free-text fields, raising live-LLM reduction to 85.1--95.6\% and author-schema reduction from 71.1\% to 90.9\%.  Median latency is 1.77\,ms.
\end{abstract}


\section{Introduction}
\label{sec:introduction}
\IEEEpubidadjcol

Consider a user who asks their LLM-powered assistant: ``What will the weather be like near my oncologist at Memorial Sloan Kettering next Thursday?''  The agent calls a weather API with \texttt{location}: \texttt{"Memorial Sloan Kettering Cancer Center, 1275 York Avenue, New York, NY 10065"}.  The weather service needs only a city-level location and a date; instead it receives the name of a cancer treatment center, revealing the user's likely health condition to a third-party operator.  This pattern is pervasive.  An agent scheduling a meeting creates a calendar event titled ``Chemotherapy Session -- Dr.\ Patel, Oncology,'' leaking a diagnosis to a calendar service.  An agent helping locate a support group searches for ``Alcoholics Anonymous meeting near 742 Evergreen Terrace, Springfield,'' exposing both addiction status and home address.  In each case, the tool functions correctly with far less data.  The agent over-shares because it optimizes for task completion, not privacy.

We call this \emph{tool call over-sharing}: the systematic inclusion of privacy-sensitive data (PSD) in tool call arguments beyond what tools require. The phenomenon is empirical, not theoretical. Across 20 everyday tasks on three production LLMs (GPT-4o, Claude 3.5 Sonnet, Llama-3.3-70B), 81--88\% of tool calls contain unnecessary PSD under default system prompts (Section~\ref{sec:motivating_study}); explicit privacy instructions cut this only to 36--76\%, and Llama barely responds to them at all (81\%$\to$76\%).  Over-sharing crosses trust boundaries to third-party services with every invocation, is invisible to users, and is uncontrolled by existing frameworks. Every agent with tool-use capability~\cite{schick2024toolformer,qin2024toollearning} is affected, and exposure scales linearly with the number of tool calls per task.

\paragraph{Why existing defenses are not enough.}
\textbf{Allow/block gating}, whether call-level (AudAgent~\cite{zheng2025audagent}, PrivacyChecker~\cite{chen2025privacyinaction}) or argument-level (Progent~\cite{shi2025progent}, which uses JSON Schema but still emits allow/block with fallback actions), cannot rewrite argument values: a weather query with ``Memorial Sloan Kettering'' either proceeds and leaks the facility or is refused, losing the task.  \textbf{Information flow control} (Fides~\cite{costa2025securingagents}, RTBAS~\cite{zhong2025rtbas}, CaMeL~\cite{debenedetti2025camel}) tracks confidentiality--integrity labels at flow granularity with capability-based allow/deny; it does not perform graduated rewriting of individual fields, and its label lattices are not designed for PSD sensitivity stratification.  \textbf{PII detection tools} (Presidio, AWS Comprehend) are token-level pattern matchers that miss \emph{implicit} PSD (``Memorial Sloan Kettering'' is not a PII pattern but implies cancer) and are schema-unaware.  \emph{No existing system performs schema-aware, argument-level data minimization by rewriting tool call fields while preserving task validity}, and none define a quantitative privacy cost metric for tool call exposure.

\paragraph{Contributions.}
We present \system{}, a middleware that intercepts LLM agent tool calls before execution, classifies PSD in their arguments, computes a quantitative privacy cost, and rewrites arguments to the minimum necessary data.  It sits between the agent and the tool without changes to the LLM, framework, or tools.  Over-sharing is an inference-time phenomenon: the helpfulness objective pushes the LLM to include all context in tool arguments regardless of safety training, and our motivating study confirms that privacy prompts still leave 36--76\% of calls leaking PSD. Intervention therefore must act on the LLM's \emph{output}.

This paper contributes:

\begin{enumerate}[leftmargin=*,nosep]
  \item A practical taxonomy of 10 PSD categories grounded in GDPR Art.~9 classifications, and a quantitative privacy cost metric $\pc(t, \mathit{args})$ combining data sensitivity, tool trust, and argument necessity.  The metric is designed for \emph{ordinal ranking} of methods: Kendall's $\tau = 1.0$ across 110 weight perturbations, so our main ranking result is invariant to parameterization.
  \item \system{}, to our knowledge the first system that applies \emph{graduated rewriting} (removal, generalization, substitution, truncation) at argument granularity while preserving task validity. Progent~\cite{shi2025progent} is also schema-aware over arguments but enforces via allow/block with fallback actions; \system{} mutates field values instead. The pipeline combines pattern+semantic PSD classification, JSON Schema necessity analysis, and the four rewriting operations.
  \item A controlled motivating study on three production LLMs (120 API calls) that establishes the empirical basis of tool call over-sharing, complementing concurrent benchmarks~\cite{trivedi2025agentdam,han2025topbench,kang2025agentleak}.
  \item A comprehensive evaluation across three agent frameworks (AutoGen, MCP, LangChain), four LLM profiles, 90 author scenarios, 50 external scenarios, and 25 real-world MCP schemas.  Live validation on 307 tool calls across GPT-4o, Claude Sonnet, and Llama-3.3-70B achieves 81.2--92.0\% privacy cost reduction at 100\% task validity (TOST $p{<}0.001$ at $\Delta{=}1.0$); on unannotated MCP schemas, 79.0\%; on Stage~C author scenarios, 73.6\% (7.3$\times$ lower residual PCS than PII Detection).
  \item AgentPrivBench, an open-source benchmark with ground-truth PSD annotations and standardized privacy metrics.
\end{enumerate}

\section{Motivating Study}
\label{sec:motivating_study}

Before designing a mitigation, we ask: \emph{do production LLMs actually over-share PSD in tool calls, and can safety training or privacy instructions prevent it?}  Prior work motivates agent privacy risks from theoretical threat models~\cite{he2024emerged,radosevich2025mcpsafety} or benchmark-only evaluations.  Concurrent work corroborates the concern: AgentDAM~\cite{trivedi2025agentdam} finds web-navigation agents are prone to unnecessary sensitive data use, TOP-Bench~\cite{han2025topbench} reports a 90\% average risk leakage rate via tool orchestration, and AgentLeak~\cite{kang2025agentleak} measures 69\% system exposure in multi-agent configurations.  We complement these with the first controlled measurement of \emph{tool call argument-level} over-sharing by production LLMs.

\paragraph{Methodology.}
We constructed 20 everyday agent requests that naturally contain personal information, the kind of tasks users routinely delegate (e.g., ``check the weather near my doctor's office,'' ``email my manager asking for a day off'').  Each prompt specifies 1--2 tools from a standard set (weather, calendar, search, email, maps) with ground-truth annotations of \emph{necessary} vs.\ \emph{unnecessary} PSD.  We queried three production LLMs (GPT-4o, Claude 3.5 Sonnet, Llama-3.3-70B) via function-calling APIs at temperature~0, under two conditions: \emph{Default} (standard helpful system prompt) and \emph{Privacy} (explicit instructions to minimize PSD with concrete examples).  This yields $3 \times 20 \times 2 = 120$ API calls, each annotated for unnecessary PSD.

\paragraph{Over-sharing is pervasive.}
Under the default prompt, \textbf{81--88\% of tool calls contain at least one unnecessary PSD item} (Table~\ref{tab:motivating}).  All three models over-share at similar rates despite differing safety training: GPT-4o (81.5\%), Claude (87.5\%), Llama (81.0\%).  In 10 of 20 scenarios, all three models unanimously over-share; in 15 of 20, at least two do.

\begin{table}[t]
\centering
\caption{Motivating study: production LLM over-sharing rates on 20 realistic agent tasks.  \emph{Over-share\%} = fraction of calls with $\geq 1$ unnecessary PSD item; \emph{Leakage\%} = fraction of possible unnecessary items actually leaked.}
\label{tab:motivating}
\footnotesize
\setlength{\tabcolsep}{3pt}
\begin{tabular}{@{}llrrr@{}}
\toprule
\textbf{Model} & \textbf{Condition} & \textbf{Over-share\%} & \textbf{Avg Unnec.} & \textbf{Leakage\%} \\
\midrule
\multirow{2}{*}{GPT-4o}
  & Default & 81.5\% & 1.00 & 40.9\% \\
  & Privacy & 36.0\% & 0.40 & 15.9\% \\
\midrule
\multirow{2}{*}{Claude 3.5 Sonnet}
  & Default & 87.5\% & 0.94 & 34.9\% \\
  & Privacy & 50.0\% & 0.56 & 20.4\% \\
\midrule
\multirow{2}{*}{Llama-3.3-70B}
  & Default & 81.0\% & 1.00 & 38.9\% \\
  & Privacy & 76.2\% & 0.86 & 33.3\% \\
\bottomrule
\end{tabular}
\end{table}

Representative examples: a weather query about a cardiology appointment produces \texttt{location}=``9500 Euclid Avenue, Cleveland OH'' with \texttt{context}=``cardiology appointment at 2pm''; the API needs only ``Cleveland, OH.''  All three models title a couples-counseling event ``Couples Counseling with Dr.\ Patricia Hoffman'' instead of a generic ``Appointment.''  When asked to email a manager about a day off, all three include ``severe anxiety and panic attacks since the layoffs.''  One model sends a user's SSN and income to an accountant.

\paragraph{Privacy instructions help some models, but not enough.}
Explicit privacy prompts reduce over-sharing for GPT-4o (81.5\%$\to$36.0\%) and Claude (87.5\%$\to$50.0\%), but have minimal effect on Llama (81.0\%$\to$76.2\%).  Even in the best case (GPT-4o with privacy prompt), over one-third of tool calls still contain unnecessary PSD.  Five scenarios resist all prompting: addiction searches, custody queries, STI clinic lookups, home-to-rehab directions, bankruptcy searches.

\paragraph{Safety training is necessary but insufficient.}
RLHF and safety fine-tuning help but cannot remove the structural incentive to include all available context. The helpfulness that makes agents effective is the same behavior that leaks PSD, which motivates middleware operating on the agent's \emph{output} rather than relying on the model to self-censor.

\section{Threat Model}
\label{sec:threat_model}

\subsection{System Model}
\label{sec:system_model}

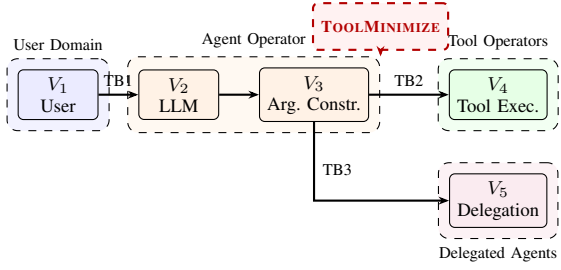
\begin{figure}[t]
  \centering
  \begin{tikzpicture}[
    node distance=0.9cm and 0.6cm,
    every node/.style={font=\footnotesize},
    stage/.style={draw, rounded corners=2pt, minimum height=0.6cm, minimum width=1.2cm, align=center, fill=white},
    trust/.style={draw, dashed, rounded corners=4pt, inner sep=4pt},
    arr/.style={-{Stealth[length=4pt]}, thick},
    scale=0.88, transform shape,
  ]
    \node[stage, fill=blue!8] (v1) {$V_1$\\User};
    \node[stage, fill=orange!10, right=of v1] (v2) {$V_2$\\LLM};
    \node[stage, fill=orange!10, right=of v2] (v3) {$V_3$\\Arg.\ Constr.};
    \node[stage, fill=green!10, right=1.2cm of v3] (v4) {$V_4$\\Tool Exec.};
    \node[stage, fill=purple!8, below=0.8cm of v4] (v5) {$V_5$\\Delegation};

    \node[draw, thick, dashed, red!70!black, fill=red!5, rounded corners=2pt,
          minimum height=0.7cm, minimum width=1.2cm,
          above=0.15cm of v3, xshift=1.0cm, yshift=0.1cm, font=\scriptsize\bfseries] (tm) {\system{}};

    \begin{scope}[on background layer]
      \node[trust, fill=blue!5, label={[font=\scriptsize]above:User Domain}] (tb1) [fit=(v1)] {};
      \node[trust, fill=orange!5, label={[font=\scriptsize]above:Agent Operator}] (tb2) [fit=(v2)(v3)] {};
      \node[trust, fill=green!5, label={[font=\scriptsize]above:Tool Operators}] (tb3) [fit=(v4)] {};
      \node[trust, fill=purple!5, label={[font=\scriptsize]below:Delegated Agents}] (tb4) [fit=(v5)] {};
    \end{scope}

    \draw[arr] (v1) -- node[above, font=\scriptsize]{TB1} (v2);
    \draw[arr] (v2) -- (v3);
    \draw[arr] (v3) -- node[above, font=\scriptsize]{TB2} (v4);
    \draw[arr] (v3) |- node[right, font=\scriptsize, pos=0.3]{TB3} (v5);

    \draw[-{Stealth[length=4pt]}, thick, red!70!black, dashed] (tm) -- ++(0,-0.45);
  \end{tikzpicture}
  \caption{LLM agent pipeline. Shaded regions denote trust domains. \system{} intercepts tool calls at the agent--tool boundary (dashed box), between argument construction ($V_3$) and tool execution ($V_4$).}
  \label{fig:pipeline}
\end{figure}

An LLM agent processes a user request through five stages:
user input ($V_1$), LLM reasoning ($V_2$), tool argument
construction ($V_3$), tool execution ($V_4$), and cross-agent
delegation ($V_5$).  Figure~\ref{fig:pipeline} illustrates the pipeline.
Three trust boundaries partition it.  The user--agent boundary ($V_1 \to V_2$) is trusted: the user chooses to interact.  The agent--tool boundary ($V_3 \to V_4$) is semi-trusted: the agent operator selects the tool, but the tool operator is a distinct entity with its own data policies; a single request may trigger 5--20 tool calls~\cite{he2024emerged}.  The agent--agent boundary ($V_2 \to V_5$) has variable trust: the user consented only to the primary agent's practices.

\system{} sits at the agent--tool boundary and intercepts every tool call after argument construction ($V_3$) but before execution ($V_4$); it requires no modifications to the LLM, framework, or tools.  For highest accuracy, tool schemas should carry \texttt{minimum\_\allowbreak{}necessary} annotations; without them, the analyzer falls back to the standard \texttt{required}/\texttt{optional} distinction.

\subsection{Adversary Model}
\label{sec:adversary_model}

We consider three adversary classes.  \emph{Semi-honest tool operators} follow the protocol but log and analyze received argument data for profiling, resale, or surveillance (the honest-but-curious model~\cite{he2024emerged,radosevich2025mcpsafety,hou2025mcplandscape}).  \emph{Network observers} passively observe tool call payloads on the network path, including via corporate proxies or API gateways that terminate TLS~\cite{errico2025securingmcp}.  \emph{Multi-tool aggregators} correlate argument data across tool calls from the same session to reconstruct user profiles~\cite{patil2025sumleaks}.

\textbf{Out of scope.}  Malicious LLM providers, adversarial users, prompt injection attacks, and training data memorization are orthogonal threats addressed by complementary systems~\cite{debenedetti2025camel,costa2025securingagents,panda2025privacy}.

\subsection{Privacy Threats}
\label{sec:privacy_threats}

We identify five privacy threats at agent--tool and agent--agent boundaries:

\medskip
\noindent\textbf{C-1: Tool call over-sharing.}
The LLM includes PSD beyond what the tool requires; this is the most pervasive threat (Section~\ref{sec:motivating_study}).  Free-text fields (queries, descriptions, notes) are the primary vectors.

\medskip
\noindent\textbf{C-2: Tool-level data exfiltration.}
A semi-honest operator extracts PSD from received arguments for profiling, resale, or surveillance.  The tool functions correctly; the operator simply retains data it legitimately receives.

\medskip
\noindent\textbf{C-3: Cross-tool aggregation.}
A multi-tool aggregator correlates PSD fragments across calls to reconstruct user profiles that no single call exposes~\cite{patil2025sumleaks}.

\medskip
\noindent\textbf{E-1: Delegation-chain privacy degradation.}
Delegation payloads typically include the full conversational context even when the subtask requires only a narrow slice, exposing PSD to agents under weaker privacy policies~\cite{schroeder2025openmultiagent,patil2025sumleaks}.

\medskip
\noindent\textbf{E-2: Context over-forwarding.}
Mirrors C-1 at the agent--agent boundary~\cite{li2025masleak}: the receiving agent may itself invoke tools, creating transitive exposure chains across multiple trust boundaries.

\subsection{Scope and Assumptions}
\label{sec:scope}

\system{} targets \textbf{C-1 (tool call over-sharing)} as its primary threat.  C-1 is the most pervasive privacy threat: it affects every tool call, crosses trust boundaries with each invocation, and is invisible to users.  By reducing PSD in arguments, \system{} also mitigates C-2 (reducing data available for exfiltration) and limits fragments available for C-3 (cross-tool aggregation).  We evaluate effectiveness against delegation threats (E-1, E-2) in Section~\ref{sec:evaluation}.

We assume: (1)~tools expose JSON Schema input schemas~\cite{hou2025mcplandscape} (tools without schemas receive conservative defaults); (2)~arguments are JSON key-value pairs; (3)~the framework exposes a hook between argument construction and execution; and (4)~tool success is observable from the response.
Training data privacy~\cite{panda2025privacy}, conversational disclosure, and prompt injection~\cite{debenedetti2025camel} are orthogonal threats.

\paragraph{What \system{} guarantees.}
Every PSD item detected and marked unnecessary is removed or generalized before transmission.  Residual exposure from classifier misses is bounded by $\sum_c (1{-}r_c)\, n_c^{\text{unnec}}\, S(c)\, E(f)$, with per-category recall $r_c$ from $0.43$ (DI) to $0.70$ (L) (Section~\ref{sec:rq3}); taking the worst case $n_c^{\text{unnec}} = n_c$ (every missed item assumed in an unnecessary field) yields $4.09$ PCS/scenario on AgentPrivBench Stage~C, vs.\ observed $0.59$ ($6.9\times$ headroom).  Adversarial encodings, out-of-taxonomy categories, necessary-field PSD retained by handlers, and rewriter failures on required fields (4/20 maps scenarios) fall outside this bound.

\section{System Design}
\label{sec:system_design}

\subsection{Pipeline Overview}
\label{sec:overview}

\begin{figure}[t]
  \centering
  \begin{tikzpicture}[
    node distance=0.35cm,
    every node/.style={font=\footnotesize},
    stage/.style={draw, rounded corners=2pt, minimum height=0.55cm, minimum width=1.3cm, align=center, thick},
    arr/.style={-{Stealth[length=4pt]}, thick},
    scale=0.85, transform shape,
  ]
    \node[stage, fill=gray!15] (agent) {Agent\\[-1pt]{\scriptsize$(t, \mathit{args})$}};
    \node[stage, fill=blue!12, right=0.35cm of agent] (classify) {Classify};
    \node[stage, fill=orange!12, right=0.35cm of classify] (score) {Score};
    \node[stage, fill=green!12, right=0.35cm of score] (analyze) {Analyze};
    \node[stage, fill=red!10, right=0.35cm of analyze] (rewrite) {Rewrite};
    \node[stage, fill=gray!15, right=0.35cm of rewrite] (tool) {Tool\\[-1pt]{\scriptsize$(t, \mathit{args}')$}};
    \draw[arr] (agent) -- (classify);
    \draw[arr] (classify) -- (score);
    \draw[arr] (score) -- (analyze);
    \draw[arr] (analyze) -- (rewrite);
    \draw[arr] (rewrite) -- (tool);
    \node[below=0.1cm of classify, font=\scriptsize, text=blue!70!black, align=center] {Pattern+Entity\\[-1pt]+Semantic};
    \node[below=0.1cm of score, font=\scriptsize, text=orange!70!black] {$\pc(t,\mathit{args})$};
    \node[below=0.1cm of analyze, font=\scriptsize, text=green!60!black, align=center] {Schema+\\[-1pt]Necessity};
    \node[below=0.1cm of rewrite, font=\scriptsize, text=red!60!black, align=center] {Remove/Gen./\\[-1pt]Sub./Trunc.};
    \node[draw, dashed, rounded corners=2pt, fill=yellow!8, font=\scriptsize,
          below=0.65cm of score, xshift=0.5cm] (schema) {Tool Schema};
    \draw[-{Stealth[length=3pt]}, dashed, gray] (schema) -- (analyze);
    \draw[-{Stealth[length=3pt]}, dashed, gray] (schema) -- (score);
  \end{tikzpicture}
  \caption{\system{} pipeline. Tool calls are intercepted after LLM generation and before execution.}
  \label{fig:architecture}
\end{figure}
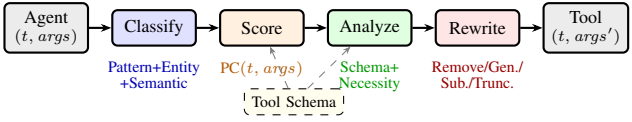

\system{} intercepts tool calls after LLM generation and before execution.  When an agent emits a call with name $t$ and arguments $\mathit{args}$, the middleware passes it through four stages (Figure~\ref{fig:architecture}): \textbf{Classify} (identify PSD items and categorize them), \textbf{Score} (compute privacy cost $\pc(t,\mathit{args})$), \textbf{Analyze} (determine field dispositions from the tool schema and necessity heuristics), and \textbf{Rewrite} (produce minimized arguments $\mathit{args}'$).  Scoring precedes analysis so every tool call is logged even when rewriting is disabled.

\subsection{PSD Taxonomy}
\label{sec:taxonomy}

We define a taxonomy of 10 PSD categories grounded in GDPR Article~9 and empirical analysis of our corpus (Table~\ref{tab:psd_taxonomy}).  We use \emph{PSD} rather than \emph{PII} because tool arguments frequently contain data that is privacy-sensitive but not personally identifying in isolation: ``Memorial Sloan Kettering'' implies a health condition, and temporal patterns reveal routines.  Sensitivity levels $S \in \{1,2,3,4\}$ are informed by GDPR regulatory tiers: Art.~9 special categories receive $S{=}4$, quasi-identifiers~\cite{sweeney2002kanonymity} receive $S{=}3$, and contextual data receive $S{\leq}2$. These levels are a technical approximation of harm, not a compliance determination.  The taxonomy is extensible; a sensitivity analysis confirming robustness to category exclusion appears in Section~\ref{sec:rq1}.

\begin{table}[t]
\centering
\caption{PSD taxonomy: 10 categories with codes, sensitivity levels, and examples.}
\label{tab:psd_taxonomy}
\footnotesize
\setlength{\tabcolsep}{3pt}
\begin{tabular}{@{}llclp{2.8cm}@{}}
\toprule
\textbf{Code} & \textbf{Category} & $S$ & \textbf{GDPR} & \textbf{Examples} \\
\midrule
DI & Direct identifiers & 4 & Art.~4(1) & SSN, email, phone, name \\
H  & Health             & 4 & Art.~9(1) & Conditions, medications \\
F  & Financial          & 4 & Art.~4(1) & Account numbers, income \\
QI & Quasi-identifiers  & 3 & Art.~4(1) & Age, ZIP, occupation, DOB \\
L  & Location           & 3 & Art.~4(1) & Street addresses, GPS \\
R  & Relational         & 3 & Art.~9(1) & Family, associations \\
B  & Behavioral         & 2 & Art.~9(1) & Habits, substance use \\
Tp & Temporal patterns  & 2 & Art.~4(1) & Schedules, routines \\
I  & Intent             & 2 & Recital~47 & Inferred goals, plans \\
Im & Implicit context   & 1 & Recital~26 & Contextual implications \\
\bottomrule
\end{tabular}
\end{table}

\subsection{Privacy Cost Metric}
\label{sec:privacy_cost}

We define the \emph{Privacy Cost} of a tool call as
\begin{equation}
\label{eq:pc}
\pc(t, \mathit{args}) = \sum_{j=1}^{|\mathcal{P}|} S(p_j) \cdot E(f) \cdot N(p_j, f)
\end{equation}
where $\mathcal{P}$ is the set of PSD items detected in $\mathit{args}$, and $f$ is the server hosting $t$.  \textbf{Sensitivity} $S(p)$ maps each PSD item to its category level (Table~\ref{tab:psd_taxonomy}).  \textbf{Exposure} $E(f)$ captures the trust relationship on four levels: \textsc{First\_Party\_Local} (0.2, on-device), \textsc{First\_Party\_Remote} (0.5, same provider), \textsc{Third\_Party\_DPA} (0.8, contracted), \textsc{Third\_Party\_No\_DPA} (1.0, uncontracted), informed by the MCP trust hierarchy~\cite{hou2025mcplandscape}.  \textbf{Necessity} $N(p, f)$ is binary: $0$ if the containing field is in the tool's \texttt{minimum\_necessary} set (derived from schema annotations), $1$ otherwise.  Only unnecessary PSD drives the score.  When schemas lack annotations, the system treats \texttt{required} fields as necessary and \texttt{optional} as unnecessary.

The metric is additive. Interaction effects (for example, a name plus a health condition is more revealing than either alone) are not modeled, but the median call has 2 categories so the effect is bounded in practice.  The base metric satisfies non-negativity, monotonicity, and additivity.  Our main acceptance result---that \system{} dominates all utility-preserving baselines---holds for \emph{any} monotone $S$ and $E$ mappings that preserve ordinal rankings: Kendall's $\tau = 1.0$ across 110 weight perturbations (Section~\ref{sec:disc-metric}).

\paragraph{Worked example.}
A weather API call (\textsc{Third\_Party\_No\_DPA}, $E{=}1.0$) with \texttt{location}~= ``Memorial Sloan Kettering Cancer Center, 1275 York Ave, NY 10065.'' The classifier detects three PSD items at non-overlapping spans: facility name (H, $S{=}4$), street address (L, $S{=}3$), ZIP (L, $S{=}3$).  The schema specifies \texttt{location} as required with \texttt{minimum\_necessary}~= [city-level], so $N{=}1$ for all three: $\pc = 4 + 3 + 3 = 10.0$.  After generalization to ``New York, NY,'' only a city-level L detection remains ($N{=}0$), yielding $\pc = 0.0$.

\subsection{Classifier, Analyzer, and Rewriter}
\label{sec:classifier}

\paragraph{Two-stage classifier.}
Stage~1 combines a \emph{PatternClassifier} (40+ regexes covering DI, H, F, L, and additional categories) with an \emph{EntityExtractor} (regex-based entity extraction mapped to PSD categories).  Stage~2 (\emph{SemanticClassifier}) catches PSD that patterns miss via sensitive facility lookup, government ID patterns, profession-based inference (e.g., ``bankruptcy attorney'' $\to$ F), and cross-field combination detection.  A base64-decoding preprocessor scans string arguments for encoded substrings; decoded spans are classified and re-mapped to their source ranges, defending against simple LLM-assisted encoding evasions (Section~\ref{sec:rq3}).  We use deterministic regex rather than statistical NER for zero external dependencies and sub-millisecond latency.  An optional spaCy NER backend (Section~\ref{sec:rq3}) raises DI recall from 0.435 to 0.739 at +26\,ms, which we recommend for deployments requiring higher coverage.

\paragraph{Analyzer.}
\label{sec:analyzer}
The analyzer decides each field's disposition in two phases.  Phase~1 (\emph{SchemaAnalyzer}) parses each tool's JSON Schema to extract required, optional, and minimum-necessary field sets; optional PSD-bearing fields receive \textsc{Remove}.  Phase~2 (\emph{SemanticNecessityAnalyzer}) examines required fields via tool-type-specific handlers: weather fields are generalized to city level, calendar titles are replaced with generic placeholders, search queries have identifiers stripped, maps queries have facility names stripped, and email fields are flagged for review.  A generic handler covers 62.2\% of scenarios with conservative rules: free-text fields with critical-sensitivity PSD are generalized; optional PSD fields are removed.  An optional \texttt{task\_context} parameter passes user-intent phrases that selectively retain consented PSD (unused in our benchmark).

\paragraph{Rewriter.}
\label{sec:rewriter}
Four operations correspond to analyzer dispositions: \emph{Removal} (delete optional unnecessary fields), \emph{Generalization} (coarsen values, e.g., ``1275 York Ave, NY 10065'' $\to$ ``New York, NY''), \emph{Substitution} (insert synthetic but schema-valid values like \texttt{user@example.com}), and \emph{Truncation} (round GPS, shorten ZIP+4 to ZIP-5).  Three failure modes are supported: \emph{fail-closed} (block the call and raise an error; default, per privacy-by-default~\cite{gdpr2016}), \emph{fail-open} (forward original, log incident; explicit opt-in), and \emph{fail-prompt} (forward with user-confirmation flag).  On initialization, the middleware flags schemas where \emph{all} fields are \texttt{minimum\_necessary} as suspicious, defending against semi-honest operators who might neutralize minimization via schema manipulation.

\subsection{Framework Integration}
\label{sec:integration}

\system{} integrates with three major agent frameworks through thin adapter layers sharing a common \texttt{ToolMinimizeMiddleware} core.  \texttt{AutoGenToolMinimize} wraps registered tool functions via a decorator or agent-level patch; \texttt{MCPToolMinimize} intercepts \texttt{tools/call} JSON-RPC requests at the MCP client transport; \texttt{LangChainToolMinimize} wraps \texttt{BaseTool} instances in a proxy.  All adapters share the same pipeline; our evaluation confirms identical privacy results across frameworks (Section~\ref{sec:rq3}).

\section{Evaluation}
\label{sec:evaluation}

We evaluate \system{} on a benchmark of 90 realistic agent scenarios addressing three research questions.
\textbf{RQ1:} How pervasive is PSD over-sharing?
\textbf{RQ2:} How does \system{} compare with existing mitigations?
\textbf{RQ3:} What is the contribution of each pipeline component?

\subsection{Experimental Setup}
\label{sec:eval-setup}

\paragraph{Benchmark.}
\emph{AgentPrivBench} contains 90 hand-curated multi-step scenarios covering healthcare (23), personal (19), finance (18), workplace (14), travel (12), and legal (4).  Each scenario specifies a user request, tool calls, and ground-truth annotations of necessary vs.\ unnecessary PSD.  Scenarios partition into \emph{tool\_call} (Stage~C, 60 scenarios, 73 tool calls), \emph{cross\_agent} (Stage~E, 15), and \emph{reasoning} (Stage~R, 15).  Annotations follow a written rubric (GDPR Art.~9 tiers, per-tool minimum-necessary analysis).
We addressed annotation-circularity concerns with two independent checks.  (i)~GPT-4o and Claude independently annotated 35 calls; both detect more PSD than our labels (5.3/7.6 vs.\ 3.4 items) at high correlation ($\rho{=}0.63, 0.79; p{<}0.001$), i.e., our labels are conservative rather than inflated.  (ii)~In Section~\ref{sec:rq3} we evaluate on 50 PrivacyLens-style external scenarios with no shared annotations, where \system{} achieves 89.1\% cost reduction.

\paragraph{LLM profiles.}
We evaluated under four simulated profiles modeled after GPT-4o, Claude~3.5, Llama-3.1-70B, and Qwen2.5-72B, each applying deterministic parameterized mutations to scenario-defined tool calls, varying optional-field inclusion, verbosity, and structure.  This isolates \system{}'s effect from LLM variance; we validate with live LLM inference on the newer GPT-4o, Claude Sonnet, and Llama-3.3-70B in Section~\ref{sec:rq3}.

\paragraph{Baselines.}
Nine methods: \emph{No Mitigation}; three \emph{Prompt} strategies (instruction $p{=}0.20$, few-shot $p{=}0.35$, system $p{=}0.40$, calibrated from \cite{li2024privacylens,mireshghallah2024confaide}); \emph{PII Detection} (Presidio-style regex + named-entity recognition (NER)); \emph{PrivacyChecker}~\cite{chen2025privacyinaction} (contextual-integrity (CI) gating that blocks any call carrying critical PSD to third-party tools; since 97\% of scenarios contain critical PSD, TCR~$= 3\%$ by design); \emph{PrivacyChecker-relaxed} (blocks only on H~+~F, not DI; TCR~$= 37\%$); \emph{AudAgent}~\cite{zheng2025audagent} (re-implementation of its blocking mode); and \system{} (rule-based backends, no LLM inference).

\paragraph{Metrics.}
\emph{PER}: fraction of calls with any PSD after rewriting.
\emph{PCS}: base $\pc(t,\mathit{args})$ averaged per scenario.
\emph{DMS} (Data Minimization Score): mean per-scenario removal rate; bounded by the fraction of unnecessary PSD (theoretical max $\approx$61.6\%).
\emph{TCR}: fraction of scenarios with schema-valid, minimum-necessary-complete rewritten arguments---\emph{argument-level} validity, not end-to-end task success (which we measure separately by execution in Section~\ref{sec:rq3} and discuss in Section~\ref{sec:disc-utility}).
Stochastic methods use $n{=}5$ runs with seeds $\{42,123,456,789,2024\}$ and report 95\% CIs; pairwise comparisons use Wilcoxon signed-rank on $n{=}60$ per-scenario PCS pairs.  Experiments ran on Intel Xeon Gold 6248R, Python 3.11, no GPU.

\subsection{RQ1: Over-Sharing Is Pervasive}
\label{sec:rq1}

Across all four LLM profiles, baseline PER averages 0.997 with 5.88 PSD items per call (${\approx}3.7$ unnecessary on average) and raw PCS 13.98.  The near-identical exposure across profiles (range $13.62$--$14.15$) supports the claim that over-sharing is structural.  Healthcare has the highest raw PCS (18.18); personal has the highest density (7.71 items/call); even legal, the lowest-density domain (2.75 items/call), transmits unnecessary PSD in every invocation.

\paragraph{Category distribution.}
69.4\% of PSD items are in critical-sensitivity categories (DI 33.0\%, H 26.6\%, F 9.8\%).  Health-related PSD appears across \emph{all} domains: agents include medical context in travel, workplace, and financial queries.  Even restricting the taxonomy to the six highest-sensitivity categories (excluding I, Im, Tp, B, 5.9\% of items), PER remains above 0.99, confirming that baseline over-sharing is not an artifact of broad taxonomy definitions.

\subsection{RQ2: Mitigation Comparison}
\label{sec:rq2}

Table~\ref{tab:rq2-comparison} compares the nine methods on 60 Stage~C scenarios.  Prompt-based methods yield modest reductions (best: 33.2\%, PER still $>$0.96).  PII Detection reduces PCS to 4.29 (27.6\% reduction) but misses non-entity PSD.  PrivacyChecker achieves PCS~$= 0.47$ at catastrophic utility loss (TCR~$= 3\%$); relaxing to block only on H~+~F raises TCR to 37\% but PCS to 3.97.  AudAgent's blocking mode (our re-implementation) acts post-hoc on detected items but does not rewrite argument fields.

\begin{table}[t]
\centering
\caption{Comparison on 60 Stage~C scenarios ($n{=}5$ runs for prompt methods). \system{} achieves the lowest PCS at TCR$=100\%$ ($p{<}0.001$ vs.\ utility-preserving baselines, Wilcoxon). Progent~\cite{shi2025progent} is bracketed by PrivacyChecker and PrivChk-relaxed (Table~\ref{tab:comparison}). $\pm$ = 95\% CI. \textbf{Bold} = best at TCR$=100\%$.}
\label{tab:rq2-comparison}
\footnotesize
\setlength{\tabcolsep}{2pt}
\begin{tabular}{@{}lccccc@{}}
\toprule
\textbf{Method} & \textbf{PER}$\downarrow$ & \textbf{PCS}$\downarrow$ & \textbf{TCR}$\uparrow$ & \textbf{DMS}$\uparrow$ & \textbf{Red.}$\uparrow$ \\
\midrule
No Mitigation      & $1.000$          & $11.68$          & $1.00$ & $0.0\%$  & $0.0\%$ \\
Prompt (instr.)    & $0.99{\pm}.01$   & $9.27{\pm}.62$   & $1.00$ & $20.8\%$ & $17.2\%$ \\
Prompt (few-shot)  & $0.98{\pm}.01$   & $7.29{\pm}.64$   & $1.00$ & $35.7\%$ & $29.8\%$ \\
Prompt (system)    & $0.97{\pm}.02$   & $6.71{\pm}.61$   & $1.00$ & $40.4\%$ & $33.2\%$ \\
PII Detection      & $0.717$          & $4.29$           & $1.00$ & $46.1\%$ & $27.6\%$ \\
PrivacyChecker     & $0.050$          & $0.47$           & $0.03$ & $95.0\%$ & $73.2\%$ \\
PrivChk-relaxed    & $0.370$          & $3.97$           & $0.37$ & $63.0\%$ & $63.9\%$ \\
AudAgent (detect)  & $1.000$          & $11.68$          & $1.00$ & $0.0\%$  & $0.0\%$ \\
\midrule
\textbf{ToolMinimize} & $\mathbf{0.567}$ & $\mathbf{0.59}$ & $\mathbf{1.00}$ & $\mathbf{62.3\%}$ & $\mathbf{73.6\%}$ \\
\bottomrule
\end{tabular}
\end{table}

\system{} achieves PCS~$=0.59$, close to PrivacyChecker's 0.47 but with 100\% task completion versus PrivacyChecker's 3\%.  The residual 0.59 corresponds to PSD that is genuinely necessary for the task (e.g., passenger names for flight booking).  On residual PCS, \system{} is $7.3\times$ lower than PII Detection and achieves 16.2\,pp higher DMS (62.3\% vs.\ 46.1\%) because it targets the items that should be removed: PII Detection misses implicit PSD and over-retains detected items in required fields.

\paragraph{Per-domain and tradeoff.}
\system{} achieves near-zero PCS in four of five Stage~C domains (finance 0.08, healthcare 0.43, personal 0.42, workplace 0.00); travel (2.00) is the exception due to named-entity identifiers that are minimum-necessary for booking.  On Stage~E (15 delegation scenarios), \system{} achieves PCS~$=0.0$ at 100\% TCR.  \system{} and PrivacyChecker form the Pareto frontier (PrivacyChecker has lower PCS but TCR~$=3\%$); all other methods are dominated ($p{<}0.001$, Wilcoxon).  These baselines pursue partly different objectives---call-level gating (AudAgent), contextual-integrity policy enforcement (PrivacyChecker), and token redaction (PII Detection)---so the comparison isolates a single axis: argument-level minimization that preserves task validity.  We do not claim general superiority over agent-security systems with different goals (e.g., information-flow integrity); \system{} is complementary to them (Section~\ref{sec:related_work}).

\subsection{RQ3: Ablation and Live Validation}
\label{sec:rq3}

\paragraph{Classifier ablation.}
Table~\ref{tab:rq3-ablation} shows incremental contributions.  Pattern matching alone detects 290 PSD items; entity extraction adds 58.6\% more (to 460).  The semantic classifier contributes only 4 additional detections.  The semantic necessity analyzer is the larger lever: schema-only necessity achieves 0.504 removal rate (234) vs.\ 0.599 full pipeline (278), generalizing PSD within required fields.

\begin{table}[t]
\centering
\caption{Ablation on 60 Stage~C scenarios.  Entity extraction adds 58.6\% more detections; semantic necessity adds 44 removals.}
\label{tab:rq3-ablation}
\small
\begin{tabular}{lrrrc}
\toprule
\textbf{Configuration} & \textbf{Det.} & \textbf{Rem.} & \textbf{Rem.\ Rate} & \textbf{TCR} \\
\midrule
Pattern-only         & 290 & 156 & 0.538 & 1.00 \\
Pattern + Entity     & 460 & 274 & 0.596 & 1.00 \\
Schema-only analysis & 464 & 234 & 0.504 & 1.00 \\
Full ToolMinimize    & 464 & 278 & 0.599 & 1.00 \\
\bottomrule
\end{tabular}
\end{table}

\paragraph{NER configuration.}
DI is the hardest category for regex classification (recall 0.435).  Adding spaCy NER (\texttt{en\_core\_web\_sm}) raises DI recall to 0.739 at +26\,ms median latency.  We report regex-only in comparison tables (for reproducibility) and enable NER for metric validation.

\paragraph{Real-world and external validation.}
On 25 MCP schemas from public servers (Shopify, GitHub, Discord, Twilio, Zoom, Jira, and 19 others) with 53 scenarios (none having \texttt{minimum\_necessary} annotations), \system{} achieves \textbf{79.0\% aggregate cost reduction}.  On 50 PrivacyLens-style external scenarios not tailored to our handlers~\cite{li2024privacylens}, \system{} achieves \textbf{89.1\% cost reduction} (PER $1.0 \to 0.16$), with TCR 60\% (external minimum-necessary definitions do not always align with schemas).  On 20 benign scenarios, detection FPR is 20\% (4/20) and rewrite FPR 10\%.

\paragraph{Production PII tool comparison and adversarial evasion.}
We ran Presidio, AWS Comprehend, and Google Cloud DLP on all 73 Stage~C calls.  \system{} detects more items per call (regex 6.30, NER 8.56) than Presidio (4.70), Comprehend (3.30), or DLP (4.42), at 50--100$\times$ lower latency; none of the three is schema-aware or rewrites arguments.  We also tested 10 PSD encoding strategies (base64, homoglyphs, leetspeak, field splitting, \ldots); with a base64 decoding preprocessor, 9/10 trigger detection. Only field splitting fully evades; compositional detection is future work.

\paragraph{Live LLM validation.}
We ran 90 scenarios through GPT-4o, Claude Sonnet, and Llama-3.3-70B, yielding 307 tool calls across 270 (scenario, model) pairs (Table~\ref{tab:live-validation}).  Schema-only \system{} achieves 81.2--92.0\% cost reduction with 100\% task completion across all three models; the content-aware extension raises this to 85.1--95.6\% (mean 91.6\%) by stripping task-irrelevant PSD from free-text fields the LLM populates.  Mann-Whitney~U finds no significant difference between real and simulated PCS ($p{=}0.085$); TOST confirms equivalence at $\Delta{=}1.0$ ($p{<}0.001$).  At tighter bounds the equivalence claim weakens: $\Delta{=}0.5$ still passes ($p{=}0.046$), but $\Delta{=}0.25$ does not ($p{=}0.23$).

\begin{table}[t]
\centering
\caption{Live LLM validation on three production LLMs (all 90~scenarios, 270~(scenario, model) pairs, 307~tool calls). TOST equivalence $p{<}0.001$ at $\Delta{=}1.0$. \textbf{Red.\%} is schema-only \system{}; \textbf{Red.\% (CA)} adds the content-aware extension. TCR is 100\% throughout. Baseline PCS gap (simulated 11.68 vs.\ real 4.10) reflects simulated profiles including all optional fields; post-minimization PCS converges (0.46 vs.\ 0.59).}
\label{tab:live-validation}
\small
\begin{tabular}{@{}lrrrr@{}}
\toprule
\textbf{Model} & \textbf{PCS$_\text{bef}$} & \textbf{PCS$_\text{aft}$} & \textbf{Red.\%} & \textbf{Red.\% (CA)} \\
\midrule
GPT-4o            & 3.95  & 0.40 & 90.0\% & 94.1\% \\
Claude Sonnet     & 5.20  & 0.42 & 92.0\% & 95.6\% \\
Llama-3.3-70B     & 3.05  & 0.57 & 81.2\% & 85.1\% \\
\midrule
\textbf{Average}  & 4.10  & 0.46 & \textbf{88.9\%} & \textbf{91.6\%} \\
Simulated (main) & 11.68 & 0.59 & 73.6\% & 90.9\% \\
\bottomrule
\end{tabular}
\end{table}

\paragraph{Content-aware extension.}
The schema-only pipeline treats any field marked \texttt{minimum\_necessary} as fully necessary and does not remove PSD from within it, so send\_email and send\_message scenarios register $\pc=0$ and 0\% reduction even when the body contains task-irrelevant disclosures (``recent job loss,'' ``unemployment benefits'').  We add an optional per-item necessity analyzer that asks an LLM (GPT-4o), for each detected PSD item, whether the specific value is essential to the user's task; items judged unnecessary are stripped from the field while the surrounding text is preserved.  On the same 90 scenarios (103 tool calls), content-aware \system{} achieves \textbf{90.9\%} mean cost reduction versus 71.1\% schema-only: all 14 send\_email and 7 send\_message calls now reach 100\% reduction, and 22 previously $\pc=0$ scenarios become nonzero.  Maps scenarios trade some reduction (100\%$\to$80\%) because the content analyzer correctly flags origin/destination addresses as task-essential.  Adding the analyzer costs one LLM call per tool invocation (${\approx}$800\,ms median, cached on replay); the schema-only path remains available as the low-latency default.

\paragraph{Framework comparison and latency.}
All three framework adapters (AutoGen, MCP, LangChain) produce identical privacy results; adapter overhead is $<$0.1\,ms.  Median latency is 1.77\,ms; P99 is 10.8\,s without timeout, inflated by deeply nested JSON ($>$5{,}000\,chars) rather than catastrophic backtracking.  A 2\,s classification timeout caps worst-case latency.  Latency scales linearly with argument size.  The sub-2\,ms schema-only path stays negligible inside multi-step agent loops: a 10-step ReAct trajectory adds ${\approx}18$\,ms of middleware latency in total, orders of magnitude below the per-step LLM inference it gates; the heavier content-aware layer (${\approx}800$\,ms/call) is applied selectively to free-text fields rather than at every step.

\paragraph{Metric validation.}
Our primary validation is ranking stability: across 110 perturbations of the $S$ and $E$ weight mappings, \system{} is the lowest-PCS utility-preserving method in 100\% of parameterizations (Kendall's $\tau=1.0$).  The comparison result in Section~\ref{sec:rq2} therefore does not depend on the specific weight choice.  As a sanity check, three LLM judges (GPT-4o, Claude Sonnet, Llama-3.3-70B) rated all 73 Stage~C calls on a harm/re-identification/sensitivity rubric (1--5); judges agree with each other (mean pairwise $\rho=0.69$) but rate a privacy-critical corpus at ceiling (range $2.2$--$5.0$, median $4.3$).  Under schema-level $N$, 18/73 calls score $\pc=0$ (all \texttt{send\_email} and \texttt{send\_message}: every detected item lies in a \texttt{minimum\_necessary} field), which yields $\rho=-0.26$ ($p=0.03$) -- a zero-variance artifact, not an inverse relationship.  Per-item $N$ judged by GPT-4o removes the collapse and recovers $\rho=+0.04$ ($p=0.74$, 95\% CI $[-0.20,+0.28]$); per-call correlation cannot climb higher against ceiling-saturated judges, but the point estimate is non-negative as required.  Discussion in Section~\ref{sec:disc-metric}.

\paragraph{End-to-end execution and linkability.}
On 20 scenarios against mock endpoints, 16/20 (80\%) produce functionally equivalent results; all four failures are maps scenarios where required addresses were removed for lack of \texttt{minimum\_necessary} annotations.  On 13 multi-call scenarios, \system{} reduces cross-call PSD linkability by 70.6\% (17$\to$5 shared items); session-level budgets would reduce this further (Section~\ref{sec:disc-future}).

\paragraph{Error analysis and recall caveat.}
Three failure modes account for residual PCS: missed DI (regex recall 0.435; NER raises to 0.739), necessary PSD retained (3/90 scenarios), and over-generalization (2 scenarios).  Headline reductions are over \emph{detected} PSD; missed items propagate unchanged, so true leak suppression is bounded above by classifier recall.

\subsection{Summary}
\label{sec:eval-summary}

Over-sharing is pervasive ($>$99\%); schema-only \system{} achieves 81.2--92.0\% on live LLM outputs (mean 88.9\%), 89.1\% on 50 external scenarios, 79.0\% on 25 unannotated MCP schemas, and 73.6\% on author schemas.  The optional content-aware extension raises live-LLM reduction to 85.1--95.6\% (mean 91.6\%) and author-schema reduction to 90.9\% by stripping task-irrelevant PSD from otherwise-necessary free-text fields.  Each pipeline stage contributes meaningfully; latency scales linearly; all three framework adapters produce identical results.

\section{Discussion}
\label{sec:discussion}

\subsection{Implications and Recommendations}
\label{sec:disc-developers}

Over-sharing is structural: 81--88\% of real LLM tool calls include unnecessary PSD. Prompting reduces this by at most 56\% (GPT-4o) and as little as 6\% (Llama). Schema-aware middleware achieves 73.6\% with full task validity. The key distinction is between schema-required fields and functionally necessary content: a weather API requires a location field but needs only city-level precision.

\paragraph{Developers.}
Deploy argument-level privacy middleware as a standard pipeline component; annotate tool schemas with \texttt{minimum\_necessary} fields (5--15 min per tool); default to \textsc{Third\_Party\_No\_DPA} for over-minimization.  Tools requiring high-precision data (maps) fail without annotations, motivating community-maintained schema registries.

\paragraph{Scaling beyond curated handlers.}
The tool-type handlers (weather, calendar, maps, \ldots) are an \emph{optimization}, not a requirement.  A single generic handler---generalize critical-sensitivity free text, drop optional PSD fields---already covers 62.2\% of scenarios, and the content-aware layer is fully tool-agnostic, judging per-item necessity from the field value and task alone.  On 25 real-world MCP schemas with \emph{no} custom handlers, \system{} still reduces cost by 79.0\% (Section~\ref{sec:rq3}).  Handlers improve precision on common tools but are not needed to onboard the long tail of APIs.

\paragraph{Protocol designers.}
MCP and other protocols currently lack privacy metadata.  We propose \emph{per-argument privacy labels}, \emph{tool-level trust declarations} in server manifests, and \emph{privacy negotiation} for data retention.  Agent-to-agent protocols should include a \texttt{required\_context} field; our Stage~E evaluation shows \system{} reduces cross-agent PCS from 12.60 to 0.00.

\subsection{Argument Validity vs.\ Task Utility}
\label{sec:disc-utility}
TCR measures \emph{argument-level} validity---rewritten arguments stay schema-valid and keep every minimum-necessary field---not end-to-end task success, and the two can diverge: generalizing a calendar event's location or stripping a maps destination can yield a valid call that no longer serves the user's intent. We therefore also report end-to-end execution against mock endpoints (Section~\ref{sec:rq3}): 16 of 20 scenarios (80\%) return functionally equivalent results, and all four failures are maps scenarios whose required addresses were removed absent \texttt{minimum\_necessary} annotations---exactly the case where the over-shared field is also the task payload. Three mechanisms narrow this gap: the content-aware layer strips task-irrelevant PSD from \emph{within} an otherwise-necessary free-text field instead of dropping the field; the optional \texttt{task\_context} parameter retains user-consented PSD; and \emph{fail-prompt} mode surfaces borderline rewrites for confirmation. Location-precise tools remain the hard case: when the destination is itself the sensitive item, routing and minimization conflict, and annotation-guided generalization (neighborhood- rather than facility-level) is the most that is safe without a user in the loop. Distinguishing the field that is structurally required from the content within it that is task-essential is the central design tension of argument-level minimization.

\subsection{Metric Validation}
\label{sec:disc-metric}

PC is designed to rank methods by how much exposure they remove, not to predict per-call judge severity. The primary validation is therefore ordinal: \emph{across 110 perturbations of the $S$ and $E$ weight mappings, the method ranking is identical (Kendall's $\tau{=}1.0$); \system{} is the best utility-preserving method in 100\% of parameterizations.}  This is the property the comparison results in Section~\ref{sec:rq2} actually depend on, and it holds regardless of the specific weight choice.

Per-call correlation with LLM judge severity is a secondary sanity check. Three LLM judges (GPT-4o, Claude Sonnet, Llama-3.3-70B) agree with each other ($\rho{=}0.69$) but rate a privacy-critical corpus near ceiling (median 4.3/5), so even a perfect metric could not produce strong rank correlation against compressed judge scores.  Under schema-level $N$ the aggregate $\rho{=}{-}0.26$ ($p{=}0.03$), driven entirely by 18/73 messaging calls receiving $\pc{=}0$ when all PSD lives in \texttt{minimum\_necessary} fields; per-item $N$ judged by an LLM removes the zero-variance collapse and recovers $\rho{=}{+}0.04$ (95\% CI $[-0.20,{+}0.28]$). The point estimate is non-negative, which is what the metric needs given saturation.  The ranking-monotonicity claim is what carries the comparison.

\subsection{Limitations, Error Analysis, and Future Work}
\label{sec:disc-limitations}

\paragraph{Limitations.}
\emph{Simulated vs.\ live}: live validation on 307 tool calls shows no significant difference (Mann-Whitney $p{=}0.085$; TOST $p{<}0.001$ at $\Delta{=}1.0$, not at $\Delta{=}0.25$).  \emph{TCR scope}: schema-validity, not end-to-end (80\%; 4/20 maps failures from missing annotations; external TCR drops to 60\%).  \emph{Ecological validity}: the 99.7\% baseline PER reflects a privacy-critical corpus; general use is lower.  \emph{User agency}: \system{} decides unilaterally; classifier targets US formats; generalization may signal concealment.  \emph{Defense}: 1/10 encodings (field splitting) still evades; 10 categories cover GDPR Art.~9 but not all special categories.  \emph{False positives}: on benign inputs detection fires on 20\% and rewriting on 10\% (Section~\ref{sec:rq3})---an over-cautious bias, mitigated by \emph{fail-prompt} confirmation, the per-item content-aware check that rejects spurious rewrites, and confidence thresholds.  Error analysis: missed DI (${\sim}40\%$ of residual), necessary PSD retained (3/90), over-generalization (2).

\paragraph{Future work.}
\label{sec:disc-future}
Adaptive NER backend for latency; session-level budgets~\cite{patil2025sumleaks}; CaMeL integration for joint confidentiality--integrity~\cite{debenedetti2025camel}; user-controlled minimization with preview UI; multimodal PSD detection.

\section{Related Work}
\label{sec:related_work}

\paragraph{Privacy in LLM systems.}
Prior work addresses training-time memorization~\cite{panda2025privacy} and conversational disclosure~\cite{mireshghallah2024confaide,li2024privacylens,li2025privaci}.  Those adversaries extract data \emph{from} the model; tool call over-sharing is the model actively \emph{sending} data to third parties.

\paragraph{Agent security and privacy.}
AudAgent~\cite{zheng2025audagent} auto-formalizes natural-language privacy policies via cross-LLM voting and blocks policy-violating actions, but its intervention is binary (allow/block) at the action level rather than field-level rewriting.  PrivacyChecker~\cite{chen2025privacyinaction} extracts the contextual-integrity (CI) five-tuple~\cite{nissenbaum2010privacy} per information flow and gates entire tool invocations; it reasons about attributes but does not mutate argument values, so it cannot distinguish a weather query with ``New York, NY'' from one with ``Memorial Sloan Kettering.''  Progent~\cite{shi2025progent} also uses JSON Schema to express fine-grained policies over tool call arguments, but enforces via allow/block with fallback actions rather than argument-level rewriting.  Das et al.~\cite{das2025beyond} audit contextual privacy with detection focus.  AgentPrint~\cite{zhang2025agentprint}, MEXTRA~\cite{wang2025unveiling}, and Patil et al.~\cite{patil2025sumleaks} address traffic fingerprinting, memory extraction, and compositional risks, respectively.  Chen et al.~\cite{chen2024privacyleakage} show users systematically underestimate privacy leakage; the invisibility of tool call arguments exacerbates this.  Concurrent benchmarks~\cite{trivedi2025agentdam,han2025topbench,kang2025agentleak} corroborate over-sharing in agent pipelines.

\paragraph{Information flow control.}
Agent IFC systems track confidentiality and integrity labels at flow granularity.  Fides~\cite{costa2025securingagents} tracks both axes with capability-based allow/deny decisions; CaMeL~\cite{debenedetti2025camel} uses capabilities to block private-data exfiltration over unauthorized flows; RTBAS~\cite{zhong2025rtbas} targets both prompt injection and privacy leakage.  All three enforce at flow granularity rather than performing graduated rewriting (removal, generalization, truncation) of individual argument fields, and their binary tainted/clean labels cannot capture the sensitivity spectrum (an email address and a cancer diagnosis carry very different exposure risks).  \system{} and these systems are complementary.

\paragraph{Data minimization and PII tools.}
GDPR Art.~5(1)(c)~\cite{gdpr2016} codifies minimization; Presidio and AWS Comprehend implement it via pattern+NER.  These tools have three limitations for agent tool calls: (i)~they miss \emph{implicit} PSD (``Memorial Sloan Kettering'' is not PII but implies cancer); (ii)~they are \emph{schema-unaware}; (iii)~they apply uniform redaction regardless of tool trust.  GAMA~\cite{yang2025gama} and DP-RAG~\cite{koga2024dprag,grislain2024ragdp} handle document-level anonymization; PAPILLON~\cite{siyan2025papillon} and Bagdasarian et al.~\cite{bagdasarian2025operationalizing} minimize at the prompt boundary; PrivWeb~\cite{yang2025privweb} protects DOM text for web agents.  Differential privacy (DP) is a poor fit for tool arguments: format constraints (email, ISO date) preclude noise.  \system{} targets a complementary boundary (tool call argument egress) with schema-based necessity and graduated rewriting (Table~\ref{tab:comparison}).

\begin{table*}[t]
\centering
\caption{Comparison with related systems.  Gran.~=~granularity; Impl.~=~implicit PSD; Sch.~=~schema-aware; Rewr.~=~rewriting; Quant.~=~metric; Util.~=~utility; Integ.~=~integrity.  \cmark~=~supported, $\circ$~=~partial, --~=~not.}
\label{tab:comparison}
\small
\setlength{\tabcolsep}{4pt}
\begin{tabular}{@{}lcccccccc@{}}
\toprule
\textbf{System} & \textbf{Gran.} & \textbf{Impl.} & \textbf{Sch.} & \textbf{Rewr.} & \textbf{Quant.} & \textbf{Util.} & \textbf{Integ.} & \textbf{Focus} \\
\midrule
AudAgent             & Flow & --    & --    & --    & --    & n/a    & --   & Det./block \\
PrivacyChecker       & Flow & $\circ$ & -- & --    & --    & Low    & --   & CI gate \\
Progent              & Arg. & --    & \cmark & -- & --    & $\circ$  & --   & Priv.\ ctrl \\
Presidio             & Tok. & --    & --    & $\circ$ & --  & $\circ$ & -- & PII redact. \\
\midrule
Fides                & Flow & --    & --    & --    & --    & n/a    & \cmark & IFC \\
RTBAS                & Flow & --    & --    & --    & --    & n/a    & \cmark & IFC \\
CaMeL                & Flow & --    & --    & --    & --    & \cmark & \cmark & IFC \\
\midrule
\textbf{ToolMinimize} & \textbf{Arg.} & \cmark & \cmark & \cmark & \cmark & \cmark & -- & \textbf{Confid.} \\
\bottomrule
\end{tabular}
\end{table*}

No existing system performs schema-aware, argument-level minimization by rewriting tool call fields while preserving task validity; \system{} fills this gap.  Integration with CaMeL or Progent for joint confidentiality--integrity is feasible (Section~\ref{sec:disc-future}).

\section{Conclusion}
\label{sec:conclusion}

Safety training cannot eliminate tool call over-sharing: the helpfulness that makes an agent useful is the same mechanism that leaks context into its tool arguments. \system{} shows that schema-aware argument rewriting reduces privacy cost by \textbf{81.2--92.0\% on 307 live tool calls} from GPT-4o, Claude Sonnet, and Llama-3.3-70B, with 100\% argument-level task validity (TOST $p{<}0.001$). The schema-only path runs at 1.77\,ms median, no LLM call required. An optional content-necessity layer that asks an LLM, per detected item, whether the value is task-essential raises live-LLM reduction to 85.1--95.6\% and author-schema reduction from 71.1\% to 90.9\% by catching PSD that lives inside otherwise-necessary free-text fields. Residual exposure is bounded by classifier recall (worst case 4.09 PCS/scenario, observed 0.59). Cross-session aggregation, adversarial field-splitting, and multimodal PSD remain open problems for agent-privacy middleware.

\bibliographystyle{IEEEtran}
\bibliography{references}

\end{document}